\documentclass[
aps,%
12pt,%
final,%
notitlepage,%
oneside,%
onecolumn,%
nobibnotes,%
nofootinbib,%
superscriptaddress,%
noshowpacs]%
{revtex4}
\usepackage{color}
\usepackage{graphicx}
\usepackage{dcolumn}
\usepackage{bm}
\usepackage{amsmath}
\usepackage{amssymb}
\usepackage{multirow}
\usepackage{xcolor}
\usepackage{float}
\usepackage{changes}
\usepackage{comment}
\allowdisplaybreaks

\def\prl{Physical Review Letters}
\def\prd{Physical Review D}

\def\jcap{JCAP }

\def\apj{ApJ}
\def\apjl{ApJL}

\def\aapr{A\&AR}

\def\mnras{MNRAS }
\def\nat{Nature}
\def\nar{New Astron. Rev.}

\def\sovast{Sov. Astron.}

\def\Mch{{\cal M}}

\def\lsim{\;\raise0.3ex\hbox{$<$\kern-0.75em\raise-1.1ex\hbox{$\sim$}}\;}
\def\gsim{\;\raise0.3ex\hbox{$>$\kern-0.75em\raise-1.1ex\hbox{$\sim$}}\;}

\def\beq#1{\begin{equation}\label{#1}}
\def\eeq{\end{equation}}
\def\beqa#1{\begin{eqnarray}\label{#1}}
\def\eeqa{\end{eqnarray}}

\def\myfrac#1#2{\left(\frac{#1}{#2}\right)}

\def\comment#1{\relax}

\begin{document}

\noindent {\it Astronomy Reports, 2023, Vol. , No. }
\bigskip\bigskip  \hrule\smallskip\hrule
\vspace{35mm}

%\noindent {\it UDK 524.42}

\title{Astrophysical appearance
of primordial black holes\footnote{Paper presented at the Fifth Zeldovich meeting, an international conference in honor of Ya. B. Zeldovich held in Yerevan, Armenia on June 12--16, 2023. Published by the recommendation of the special editors: R. Ruffini, N. Sahakyan and G. V. Vereshchagin.}}

\author{\bf \copyright $\:$  2023.
\quad \firstname{K.~A.}~\surname{Postnov}}%
\email{pk@sai.msu.ru}
\author{\bf \firstname{A.~G.}~\surname{Kuranov}}
\author{\bf \firstname{N.~A.}~\surname{Mitichkin}}
\affiliation{ Sternberg Astronomical Institute, M.V. Lomonosov Moscow State University, Universitetskij pr. 13, 119234, Moscow, Russia}%

%\author{\bf \firstname{A.~G.}~\surname{Kuranov}}
%\affiliation{Second affiliation}
%\author{\bf \firstname{N.~A.}~\surname{Mitichkin}}
%\affiliation{Second affiliation}

\begin{abstract}
Interest to astrophysical evidence for primordial black holes (PBHs) formed in the early Universe from initial cosmological perturbations has increased after the discovery of coalescing binary black holes with masses more than dozen solar ones by gravitational-wave (GW) observatories. We briefly discuss  
increasing evidence that PBHs can provide some fraction of detected merging binary BHs and can be related to an isotropic stochastic GW background recently discovered by pulsar timing arrays.
We focus on PBHs with log-normal mass spectrum originated from isocurvature perturbations 
in the modified Affleck-Dine baryogenesis scenario by Dolgov and Silk (1993). We show that almost equal populations of astrophysical binary BHs from massive binary evolution and binary PBHs with log-normal mass spectrum can describe both the observed chirp mass distribution and effective spin -- mass ratio anti-correlation of the LVK binary BHs.
\centerline{\footnotesize Received: ;$\;$
Revised: ;$\;$ Accepted: .}\bigskip\bigskip\bigskip

\end{abstract}

\maketitle

\section{Introduction}

Primordial BHs can be formed in the early radiation-dominated Universe from cosmological perturbations \cite{1967SvA....10..602Z,1974MNRAS.168..399C}.
An overdensity $\delta \rho/\rho$, upon re-entering the cosmological horizon, stops expanding and recollapses. It struggles against the pressure, and to form a BH, its size should exceed the Jeans length, which sets a critical (model-dependent) overdensity amplitude $\delta_c\sim w$ at the radiation-dominated stage with equation of state $p=w\epsilon$ \cite{1974MNRAS.168..399C}.
Thus, the mass of a PBH should be about the mass of cosmological horizon $M_\mathrm{hor} \approx (2.2 \times 10^5 M_\odot)(t/[\mathrm{s}])\approx 8 M_\odot (100 \,{\rm MeV}/ T)^2$ where $T$ is the temperature.
At the present time,  the PBH masses can span a wide range from $\sim 10^{14}$~g to billion solar masses. There are a lot of astrophysical constraints on the PBH abundances with different masses usually expressed in terms of the PBH fraction in cosmological cold dark matter density $f_{pbh}=\Omega_{pbh}/\Omega_{dm}$ \cite{2020ARNPS..70..355C}; see, e.g., the discussion of the extended PBH mass spectrum related to the thermal history at radiation-dominated stage in \cite{2023arXiv230603903C} and references therein. 

Still, the problem of PBH creation remains open, and there are different PBH formation mechanisms \cite{2020ARNPS..70..355C}. Here we will focus on the  
PBHs with log-normal mass spectrum predicted by a modified Affleck-Dine baryogenesis scenario \cite{1993PhRvD..47.4244D,2009NuPhB.807..229D} (ADD model below). In this model, at the end of the inflationary stage, a scalar field charged under baryon number ends up rotating in its inner space toward different directions depending on its location in real space to form domains (isocurvature fluctuations) with high baryon (antibaryon) charge. During the QCD phase transition ($T_\mathrm{QCD}\sim 100$~MeV, $t\sim 10^{-5}$~s), the high-density regions decouple from the cosmological expansion and produce PBHs with log-normal mass spectrum: 
\begin{equation}
    \frac{dn}{dM}=\mu^2\exp\left[-\gamma\ln^2\myfrac{M}{M_0}\right]
\end{equation}
($\mu$ is the normalization constant, $\gamma$ and $M_0$ are free parameters). This initial PBH mass spectrum was found to be appropriate to fit the entire chirp mass distribution of LVK BH binaries \cite{2019JCAP...02..018R,2020JCAP...12..017D,2022arXiv221016094L}. In addition, the ADD model naturally leads to a non-standard nucleosynthesis in high-baryon density regions which may give rise to the appearance of large matter-antimatter domains and the formation of antistars in the Universe \cite{2007NuPhB.784..132B,2015PhRvD..92b3516B,Bykov_2023}. Intermediate-mass PBHs with $M\sim 10^4 M_\odot$ formed by the ADD mechanism can also serve as natural seeds for the formation of early supermassive BHs in galactic nuclei \cite{2016JCAP...11..036B}.

Here we briefly discuss the possible signs of appearance of a binary PBH population in the existing gravitational wave observations of binary BH coalescences (LVK GWTC-3 catalog, \cite{2021_GWTC3binaries}). Specifically, we show that it is possible to describe the currently observed chirp mass distribution of coalescing binary BH by two almost equal populations: one (astrophysical) emerged from the evolution of massive binary systems in galaxies \cite{2019MNRAS.483.3288P}, with chirp masses \footnote{The chirp mass of a coalescing binary is the combination of masses of the binary components $M_1$ and $M_2$: $\Mch=(M_1M_2)^{3/5}/(M_1+M_2)^{1/5}$, which can be  determined the most accurately from the inspiraling GW signal.} up to $\Mch\sim 30 M_\odot$, and another (PBH) due to mergings of binary PBH with log-normal mass spectrum with a more extended chirp mass range. We also show that the possible anti-correlation of the binary mass ratio $q=M_2/M_1\le 1$ with the effective spin $\chi_\mathrm{eff}$of coalescing BH binaries  \footnote{The effective spin of a coalescing binary BH with individual dimensionless spins $a_1$ and $a_2$ is the mass-weighted projection of individual spins on the orbital angular momentum  prior to the merging $\chi_{\mathrm{eff}}=(M_1a_1\cos\theta_1+M_2a_2\cos\theta_2)/(M_1+M_2)$.} \cite{2022MNRAS.tmp.2764A,2023arXiv230715278A} can be explained by the assumed two binary BH populations taking into account accretion-induced effective spin evolution of binary PBHs \cite{2019JCAP...06..044P}. We also briefly discuss the possible relation of the recently reported PTA stochastic GW backgrounds \cite{2023ApJ...951L...8A,2023arXiv230616214A} with stellar-mass PBHs.

\section{Binary BH populations}

\subsection{Astrophysical binary BHs}

There is a growing literature on the possible astrophysical formation channels of the observed LVK coalescing binary BHs (see, e.g., \cite{2022LRR....25....1M} and references therein).  Main astrophysical channels assume the formation of coalescing binary BHs, in particular, from the evolution of field massive binaries in galaxies at different redshifts with different stellar metallicities \cite{1993MNRAS.260..675T,1997AstL...23..492L,2016Natur.534..512B,2017MNRAS.468.5020I}, from the dynamical capture in dense stellar systems \cite{2015PhRvL.115e1101R,2016ApJ...824L...8R}, and from the binary evolution in dense surroundings in AGNs \cite{2017ApJ...835..165B}.   

Astrophysical scenarios heavily rely on the (still uncertain) BH formation mechanisms from core collapse of massive stars (e.g. \cite{2012ApJ...749...91F,2023arXiv230805798B}). 
Here we will restrict ourselves by considering the simplest possible case of binary BH formation via massive binary field evolution considered in \cite{2019MNRAS.483.3288P} aimed at exploring the effective spins and masses of the coalescing binary BHs. In this model, we studied the BH formation from stellar evolution in two distinct cases: (1) the mass of a nascent BH from the collapse of stellar core with mass $M_{CO}$  is $M_{BH}=0.9M_{CO}$ ('model CO'), (2) the BH mass is adopted according to the model \cite{2012ApJ...749...91F} ('model BH'). In addition, the evolution of the stellar core rotation in the course of stellar evolution, tidal interactions between the binary components, and possible fallback accretion onto the BH from the stellar envelope were taken into account. 
We ignored the possible high kicks acquired by low-mass BHs in some models (e.g.,  \cite{2023arXiv230805798B}).
We applied the standard binary evolution scenario  \cite{2014LRR....17....3P} with the common envelope (CE) stage treatment using the CE efficiency parameter $\alpha_{CE}$ \cite{2013A&ARv..21...59I}. A population synthesis of binary BHs with taking into account the dependence of the stellar metallicity and star formation rate in galaxies on redshift enabled us to calculate the expected chirp mass, mass ratios and effective spin distributions of coalescing BH binaries \cite{2019MNRAS.483.3288P}.

\subsection{Primordial binary BHs}

Binary PBHs can be formed in the early Universe in many different ways \cite{1997ApJ...487L.139N,1998PhRvD..58f3003I,2017PhRvD..96l3523A,2019JCAP...02..018R}. Stellar-mass binary PBHs remain attractive sources of the observed LVK binary BH coalescences and can explain the inferred binary BH coalescence rate $\sim 18 - 44$ Gpc$^{-3}$~yr$^{-1}$ \cite{2021_GWTC3binaries} with very modest requirements on the cosmological PBH abundance $f_{pbh}\sim 10^{-3}$ in the mass rage $10-100 M_\odot$ \cite{2016JCAP...11..036B,2016PhRvL.116t1301B,2016PhRvL.117f1101S,2017PhRvD..96l3523A}. Our calculations of the binary PBH coalescence rate \cite{2023arXiv230206981P} with log-normal mass spectrum show that it is very weakly dependent on the assumed binary PBH formation scenario, although the possible PBH clustering can increase the required PBH cosmological fraction $f_{pbh}$ \cite{2023Symm...15..637E}.

\section{Two populations of coalescing binary black holes}

\begin{figure}
\begin{center}
\includegraphics[width=0.49\textwidth]{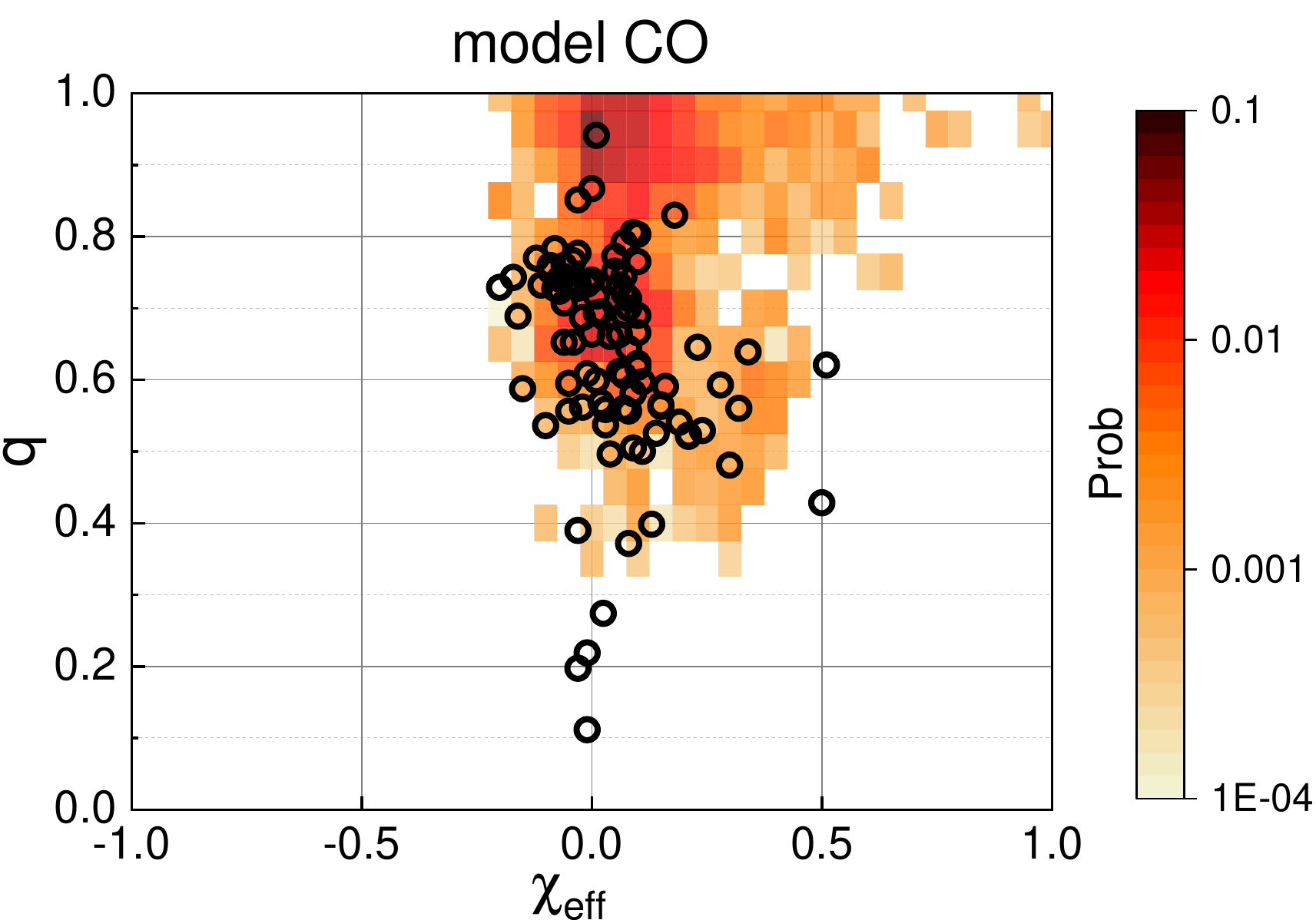}
\includegraphics[width=0.49\textwidth]{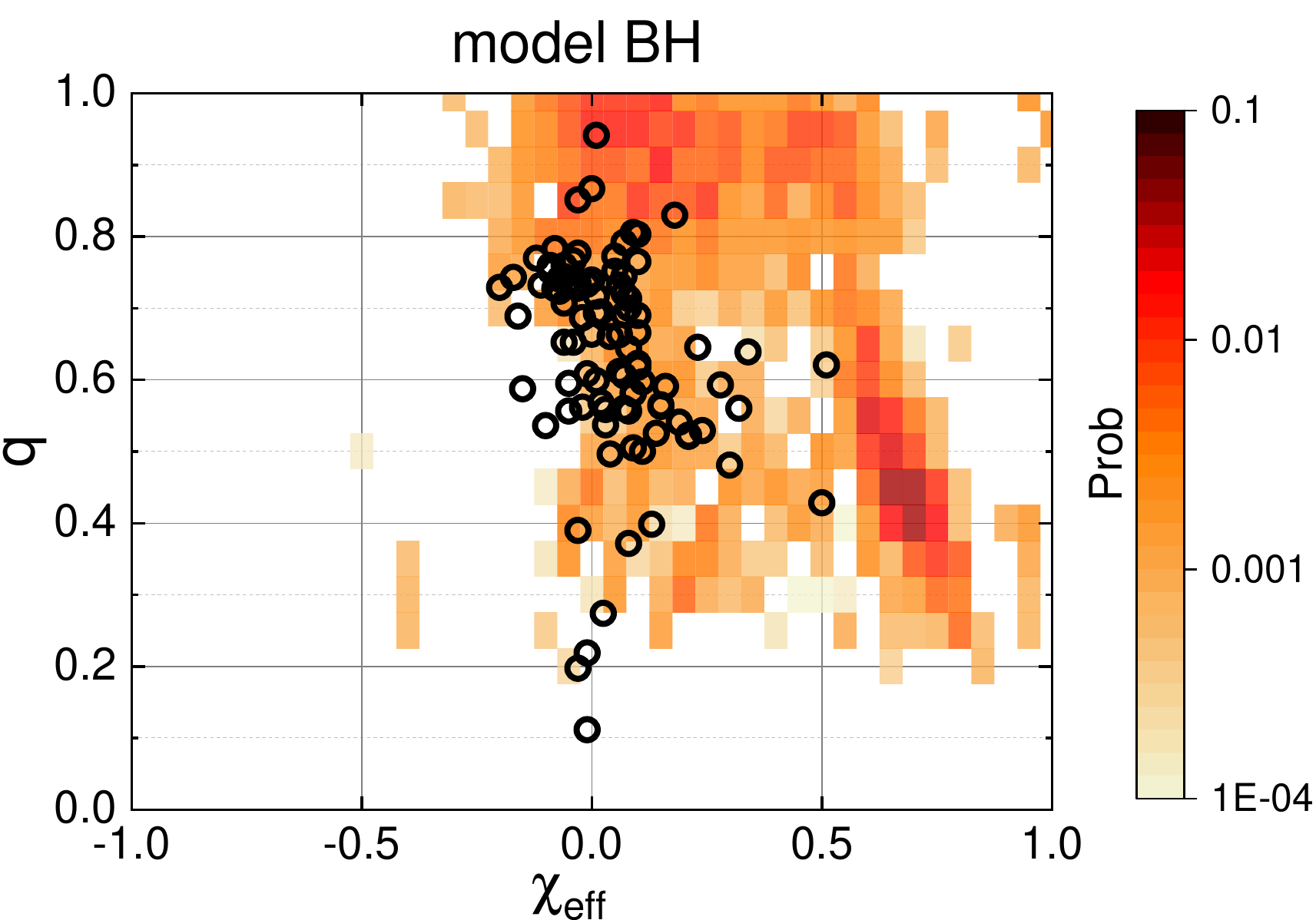}
\caption{
The mass ratio $q=m_2/m_1$ and effective spin $\chi_{eff}$ of merging binary BHs from the GWTC-3 \cite{2021arXiv211103606T} catalogue (open circles, errors are not shown) overplotted with the model normalized distribution of astrophysical merging binary BHs from the model 
\cite{2019MNRAS.483.3288P} (color grade to the right). Left panel: BH formation model with $M_{BH}=0.9 M_{CO}$ ('model CO'). Right panel: BH formation model according to \cite{2012ApJ...749...91F} ('model BH').
\label{f:astro} }
\end{center}
\end{figure}

 In Fig. \ref{f:astro}, we plot the normalized distribution of the  mass ratio $q=M_2/M_1\le 1$ vs effective spin $\chi_{eff}$  of the synthesized astrophysical population of coalescing binary BHs (colour grade to the right) overplotted with the observed data from the GWTC-3 LVK catalogue (open circles, errors are not shown) \cite{2021arXiv211103606T} for two models of BH formation from core collapses of massive stars ('model CO' and 'model BH') and standard binary evolution in galaxies with account of the stellar metallicity and star formation rate dependence on redshift \cite{2019MNRAS.483.3288P}. The standard value of the CE efficiency parameter $\alpha_{CE}=1$ is assumed. Clearly, the simplest assumption  on the BH masses from stellar core collapses with $M_{BH}=0.9M_{CO}$ (left panel) is in better agreement with GWTC-3 data, although the model predicts an overproduction of equal-mass binaries ($q\sim 1$) with a wide spread of the effective spins. 

The PBH formation mechanisms at the radiation-dominated stage predict low PBH initial spins \cite{2019JCAP...05..018D,2020JCAP...03..017M,2021PhRvD.104h3018C} (see, however, \cite{2023arXiv230611810D}). Still, the effective spin of a spinless binary PBH with high eccentricity can increase due to accretion from the ambient medium \cite{2019JCAP...06..044P}, with the highest effective spins acquired in binaries with low $q$.
In Fig. 2 we show the mass ratio $q=M_2/M_1$ and effective spin $\chi_{eff}$ of merging binary BHs from the GWTC-3 \cite{2021arXiv211103606T} catalogue (black filled circles with red error bars) and the expected $\chi_{eff}-q$ relations due to accretion from the interstellar matter onto a binary PBH with the components mass ratio $q$ and initial eccentricity $e=0.99$ in the model \cite{2019JCAP...06..044P} (coloured curves). Different colours of the model curves are used for different chirp masses (in $M_\odot$) and surrounding density (in [g~cm$^{-3}$]). It is seen that the possible effective spin -- mass ratio anti-correlation \cite{2022MNRAS.tmp.2764A,2023arXiv230715278A} can be explained by PBH binaries with high chirp masses $\Mch>10 M_\odot$, while negative effective spins can appear in astrophysical binary BHs (see Fig. \ref{f:astro}, left panel). 

In Fig. \ref{f:chirpM} we present the distribution function of chirp masses of coalescing binary BHs in two adopted astrophysical models (green and black dashed curves) and for coalescing PBHs with log-normal mass spectrum (orange and magenta dashed curves).  The PBH mass fraction needed to explain the observed coalescence rate of the LVK binary BHs is $f_{pbh}\sim 10^{-3}$. Details of calculation of the binary PBH chirp mass distribution can be found in our previous paper \cite{2023arXiv230206981P}. We see that taking almost equal fractions of both populations ($x_{abh}=0.47$, $x_{pbh}=0.53$) with PBH parameters $M_c=33 M_\odot$ and $\gamma=10$ can describe the chirp mass distribution of coalescing binary BHs from the GWTC-3 catalogue \cite{2021arXiv211103606T}. 
Note that the large value of the parameter $\gamma\sim 10$ means a fairly narrow log-normal distribution around $M_0$, while the possible fit of the overall chirp mass range of merging binary BHs in this model requires a more natural value $\gamma\sim 1$ \cite{2020JCAP...12..017D}. 
It will be interesting to test whether the apparent deficit of coalescing BH binaries with $\Mch\sim 20 M_\odot$ will persist in the ongoing O4 LVK run. 

\section{Stellar-mass PBHs in the stochastic GW background}

The discovery of an isotropic stochastic GW background at nano-Hertz frequencies was recently announced by pulsar timing arrays NANOGrav \cite{2023ApJ...951L...8A}, EPTA/InPTA \cite{2023arXiv230616214A}, PPTA \cite{2023ApJ...951L...6R} and CPTA \cite{2023RAA....23g5024X} collaborations. Such a background can be produced by a collection of unresolved supermassive BH binaries, although it can be also explained by different cosmological sources, including first-order phase transitions, cosmic strings, domain walls, and second-order GWs produced from primordial scalar curvature perturbations  \cite{2023ApJ...951L..11A,2023arXiv230616227A}. In the last case, the production of PBHs associated with density perturbations is expected \cite{2010PhRvD..81b3517B} in the mass range corresponding to the scale $\lambda=2\pi/k$ at the perturbation's cosmological horizon crossing
$k=aH$:
\begin{equation}
    M(k)\approx 30 M_\odot \left(\frac{10.75}{g_*}\right)^{-1/6}\left(\frac{3\times 10^5\rm Mpc^{-1}}{k}\right)^{2}\,.
\end{equation}
($g_*$ is the number of relativistic degrees of freedom).  The comoving scale at the horizon crossing is related to the frequency as $f\approx 1.55[\mathrm{nHz}] (k/10^6 \mathrm{Mpc}^{-1})$. The primordial origin of the detected nano-Hz stochastic GW background requires a non-standard inflationary spectrum of perturbations strongly enhanced at small scales $k\sim 10^6 \mathrm{Mpc}^{-1}$. Such peaks are possible in some non-standard inflationary scenarios (e.g., \cite{2020JCAP...08..001B}). PBHs from such perturbations can correspond to the observed enhancement in the chirp mass distribution of coalescing binary BHs at 30 $M_\odot$ discussed above and explain the observed merging rate of LVK BH binaries with modest requirements on their abundance $f_{pbh}$. 

\section{Conclusions}

Presently, the existence of PBHs in a wide mass range is consistent with numerous astrophysical observations \cite{2023arXiv230603903C}. Here we have shown that a particular PBH formation model predicting stellar-mass PBHs with log-normal mass spectrum (1) \cite{1993PhRvD..47.4244D} remains viable to describe the observed peak at $\sim 30 M_\odot$ appeared in the chirp mass distribution of coalescing binary BHs in the published LVK data \cite{2021arXiv211103606T}, as an addition to the astrophysical binary BHs arising from massive binary star evolution \cite{2019MNRAS.483.3288P} that can explain the chirp mass distribution at smaller $\Mch\sim 10 M_\odot$ (see Fig. 3). The effective spins of the observed coalescing binary BHs and possible effective spin -- mass ratio anti-correlation \cite{2023arXiv230715278A} also can be matched with the expectations from two populations of coalescing binary BHs (Fig. 1, 2). 

One of the direct pointings to the primordial formation of LVK BH binaries could be detection of light ($<2 M_\odot$) BHs among coalescing binaries in the LVK data. However, no reliable candidates have been found so far \cite{2023arXiv230816173W}. 

The extended PBH log-normal mass spectrum (1) remains also attractive to provide massive ($\sim 10^3-10^4 M_\odot$) seeds for early SMBH growth \cite{2016JCAP...11..036B}, which is a highly topical issue of the  modern astrophysics \cite{2022NewAR..9401642M}. In this model, the sufficient amount of massive seeds can be produced for the parameters of the mass spectrum $M_0\sim 20 M_\odot$ and $\gamma\sim 1$, which describes well the overall chirp mass distribution in the LVK data \cite{2020JCAP...12..017D}. However, a narrow distribution with $\gamma\sim 10$ which follows from the GWTC-3 data fit (Fig. 3 and \cite{2023arXiv230206981P}) cannot source the required density of SMBH seeds.
Increase in the statistics of the coalescing binary BHs during the ongoing O4 LVK run will show whether the peaks at the chirp mass distribution \cite{2021arXiv211103606T}, possibly corresponding to two populations of the coalescing binary BHs \cite{2022PhRvD.106l3526F,2023arXiv230206981P}, persists. 

We conclude that after more than half a century,  more and more astrophysical and cosmological evidences arise supporting the existence in nature  of PBHs  predicted in the 1970s by Zeldovich and Novikov \cite{1967SvA....10..602Z} and Carr and Hawking \cite{1974MNRAS.168..399C}. 
PBHs not only can provide a significant fraction of cosmological cold dark matter, but possibly have already been detected in the ongoing GW ground-based experiments and show up in the stochastic GW background. They remain viable candidates for the massive seeds for early SMBH growth. Following L. Wittgenstein, "The problems are solved, not by coming up with new discoveries, but by assembling what we have long been familiar with" \cite{Wittgenstein}. Do we agree?

\begin{acknowledgments}

KP thanks the organizers of the 5th Zeldovich meeting in Yerevan for wonderful organization of the conference and hospitality.
This work has made use of NASA ADS system. 
\end{acknowledgments}

\section*{Funding}
KAP and NAM acknowledge the support from the Russian Science Foundation grant  23-42-00055. The work of NAM is also supported  by the BASIS Foundation through grant 22-2-10-2-1.
The work of AGK is partially supported by Russian Science Foundation grant 21-12-00141 (calculation of chirp masses and effective spins of coalescing astrophysical binary black holes).

\clearpage
\section*{REFERENCES}
%\centerline{ÑÏÈÑÎÊ ËÈÒÅÐÀÒÓÐÛ}
%\bibliographystyle{apsrev}
%\bibliography{GW1}

%\begin{thebibliography}{100}

\clearpage

% Figure captions

{\bf Figure captions to Postnov, Kuranov and Mitichkin}
\bigskip\bigskip

Fig.~1.~The mass ratio $q=m_2/m_1$ and effective spin $\chi_{eff}$ of merging binary BHs from the GWTC-3 \cite{2021arXiv211103606T} catalogue (open circles, errors are not shown) overplotted with the model normalized distribution of astrophysical merging binary BHs from the model 
\cite{2019MNRAS.483.3288P} (color grade to the right). Left panel: BH formation model with $M_{BH}=0.9 M_{CO}$ ('model CO'). Right panel: BH formation model according to \cite{2012ApJ...749...91F} ('model BH').

\bigskip

Fig.~2.~The mass ratio $q=M_2/M_1$ and effective spin $\chi_{eff}$ of merging binary BHs from the GWTC-3 \cite{2021arXiv211103606T} catalogue (black filled circles with red error bars). The colored curves show $\chi_{eff}-q$ relations due to accretion from the interstellar matter onto a binary PBH with component mass ratio $q$ and initial eccentricity $e=0.99$ in the model \cite{2019JCAP...06..044P}. Different colours of the model curves are for different chirp masses (in $M_\odot$) and surrounding density (in [g~cm$^{-3}$]). 

\bigskip

Fig.~3.~The observed (blue step-like curve)  distribution function of the chirp-masses of coalescing binary BHs from the LVK GWTC-3 catalogue \cite{2021arXiv211103606T}. Green and black dashed curve: astrophysical BH formation 'model CO' and 'model BH', respectively, with fraction $x_{abh}=0.47$ in the total distribution. Magenta and orange dashed curves: merging binary PBH chirp mass distribution in the binary PBH formation model \cite{1998PhRvD..58f3003I} and \cite{2019JCAP...02..018R}, respectively, for log-normal PBH mass function with parameters  $M_c=33 M_\odot$ and $\gamma=10$.  The red curve shows the total chirp mass distribution with astrophysical ('model CO' with $\alpha_{CE}=1$) and PBH populations of coalescing BH binaries ($x_{abh}=0.47$, $x_{pbh}=0.53$). The combined distribution (red) fits the empirical distribution function (blue) at 90\% CL according to the modified KS-test. See \cite{2023arXiv230206981P} for more detail. 

\clearpage

%Figures

\begin{figure}
\begin{center}
\includegraphics[width=\textwidth]{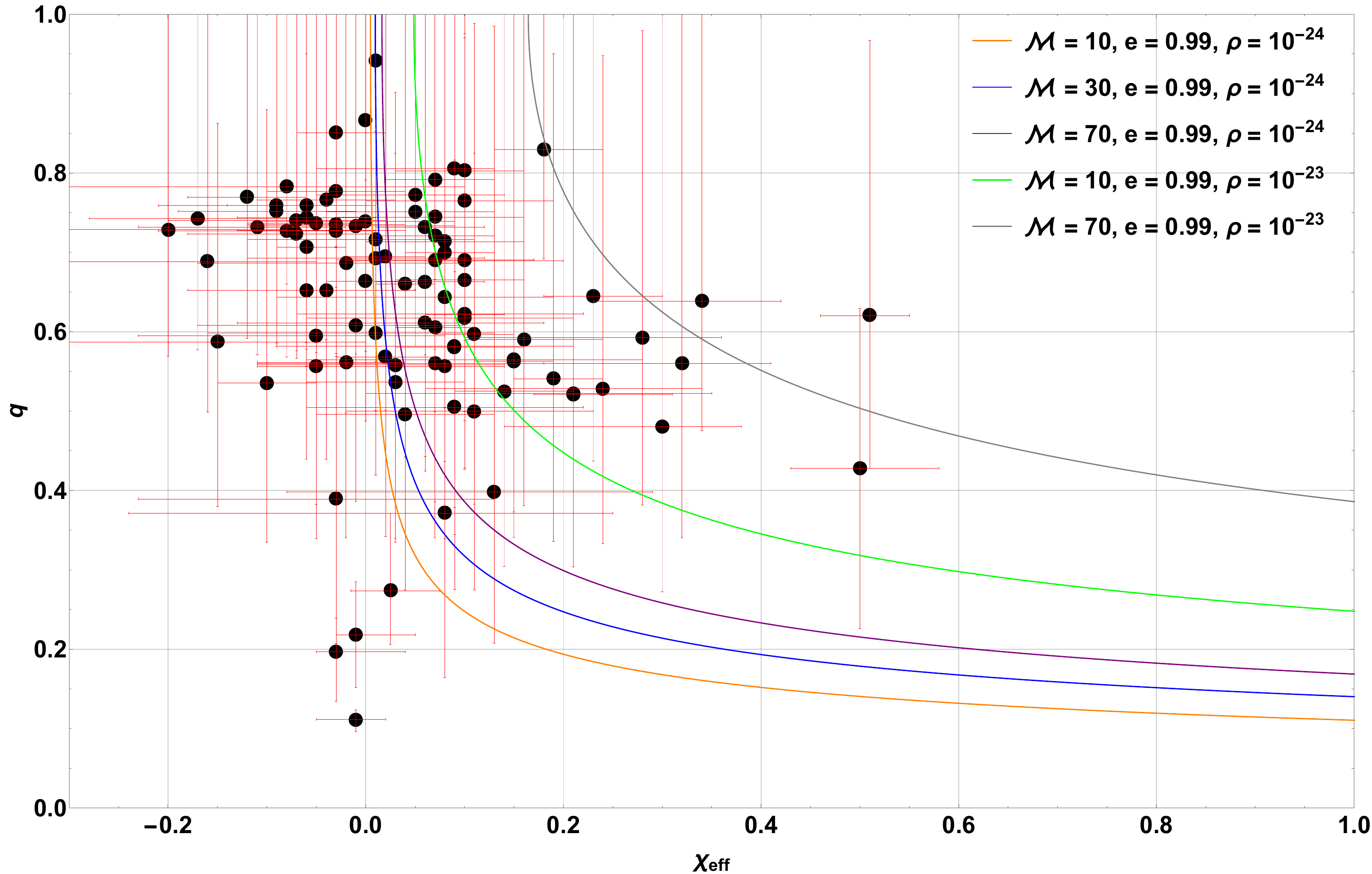}
\caption{The mass ratio $q=M_2/M_1$ and effective spin $\chi_{eff}$ of merging binary BHs from the GWTC-3 \cite{2021arXiv211103606T} catalogue (black filled circles with red error bars). The colored curves show $\chi_{eff}-q$ relations due to accretion from the interstellar matter onto a binary PBH with component mass ratio $q$ and initial eccentricity $e=0.99$ in the model \cite{2019JCAP...06..044P}. Different colours of the model curves are for different chirp masses (in $M_\odot$) and surrounding density (in [g~cm$^{-3}$]). 
}
\label{f:pbh}
\end{center}
\end{figure}

\begin{figure}
\begin{center}
\includegraphics[width=\textwidth]{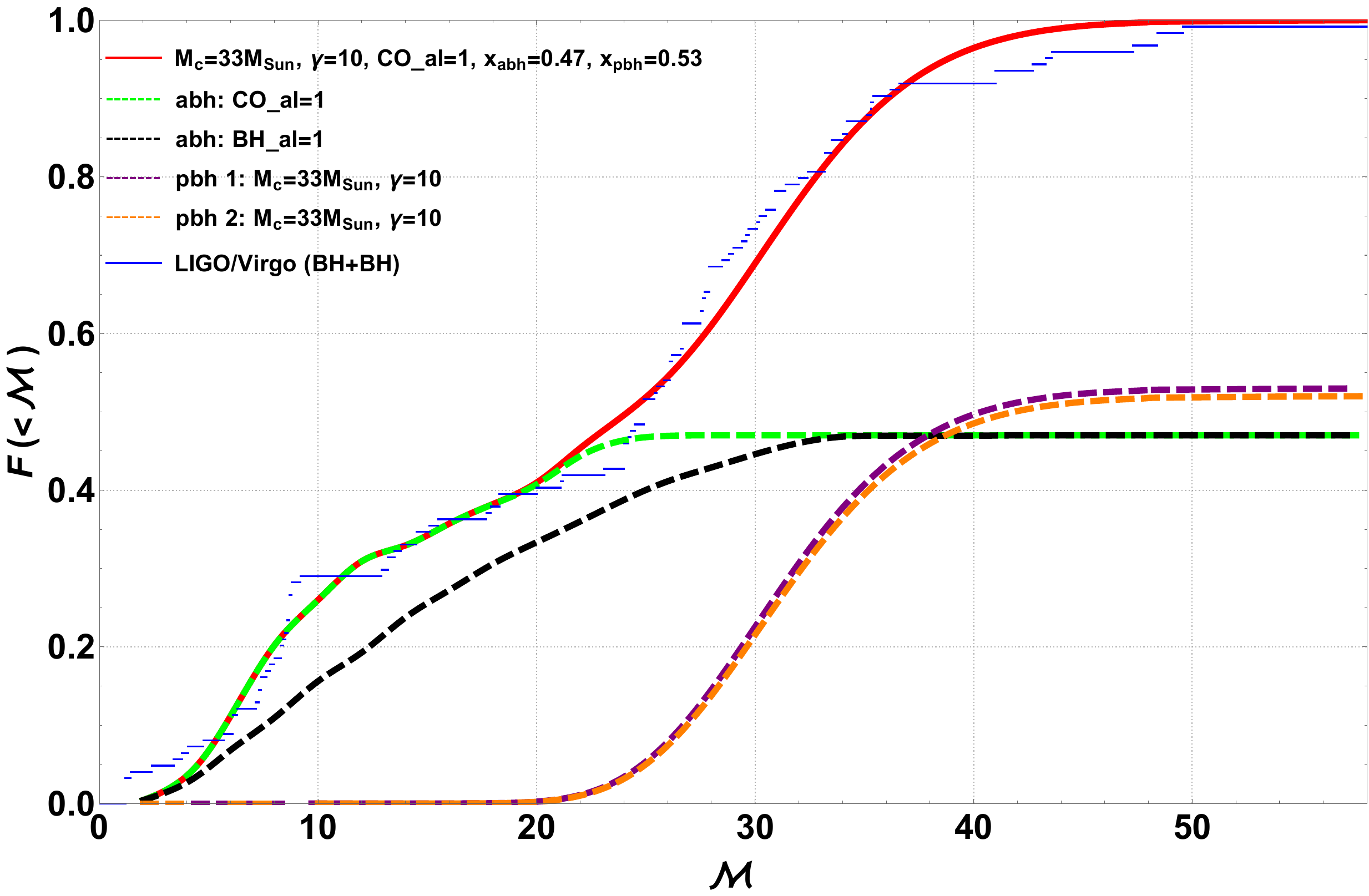}
\caption{The observed (blue step-like curve)  distribution function of the chirp-masses of coalescing binary BHs from the LVK GWTC-3 catalogue \cite{2021arXiv211103606T}. Green and black dashed curve: astrophysical BH formation 'model CO' and 'model BH', respectively, with fraction $x_{abh}=0.47$ in the total distribution. Magenta and orange dashed curves: merging binary PBH chirp mass distribution in the binary PBH formation model \cite{1998PhRvD..58f3003I} and \cite{2019JCAP...02..018R}, respectively, for log-normal PBH mass function with parameters  $M_c=33 M_\odot$ and $\gamma=10$.  The red curve shows the total chirp mass distribution with astrophysical ('model CO' with $\alpha_{CE}=1$) and PBH populations of coalescing BH binaries ($x_{abh}=0.47$, $x_{pbh}=0.53$). The combined distribution (red) fits the empirical distribution function (blue) at 90\% CL according to the modified KS-test. See
\cite{2023arXiv230206981P} for more detail. 
\label{f:chirpM} }
\end{center}
\end{figure}

\end{document}